\documentclass[lettersize,journal]{IEEEtran}
\usepackage{amsmath,amsfonts}
\usepackage{algorithmic,color}
\usepackage{array}
\usepackage[caption=false,font=normalsize,labelfont=sf,textfont=sf]{subfig}
\usepackage{textcomp}
\usepackage{stfloats}
\usepackage{url}
\usepackage{verbatim}
\usepackage{graphicx}
\usepackage{multirow}
\usepackage{booktabs}
\usepackage{pifont}
\newcommand{\cmark}{\ding{51}}%
\newcommand{\xmark}{\ding{55}}%
\usepackage{cite}
\usepackage{pifont}

\def\BibTeX{{\rm B\kern-.05em{\sc i\kern-.025em b}\kern-.08em
    T\kern-.1667em\lower.7ex\hbox{E}\kern-.125emX}}
\usepackage{balance}
\begin{document}

\title{Joint Training of Speech Enhancement and Self-supervised Model for Noise-robust ASR}

\author{Qiu-Shi Zhu, Jie Zhang, Zi-Qiang Zhang, Li-Rong Dai
\thanks{Manuscript created May, 2022; This work was supported by the National Natural Science Foundation of China (No. 62101523),
Fundamental Research Funds for the Central Universities and the Leading Plan of CAS (XDC08010200). (Corresponding author: Jie Zhang.)

The authors are with the National Engineering Research Center of Speech and Language Information Processing (NERC-SLIP), University of Science and Technology of China (USTC), Hefei 230026, China (e-mail: qszhu@mail.ustc.edu.cn; jzhang6@ustc.edu.cn; zz12375@mail.ustc.edu.cn; lrdai@ustc.edu.cn).}}

\markboth{IEEE/ACM TRANSACTIONS ON AUDIO, SPEECH, AND LANGUAGE PROCESSING}%
{How to Use the IEEEtran \LaTeX \ Templates}

\maketitle

\begin{abstract}
Speech enhancement (SE) is usually required as a front end  to improve the speech quality in noisy environments, while the enhanced speech might not be optimal for automatic speech recognition (ASR) systems due to speech distortion. On the other hand, it was shown that self-supervised pre-training enables the utilization of a large amount of unlabeled noisy data, which is rather beneficial for the noise robustness of ASR. However, the potential of the (optimal) integration of SE and self-supervised pre-training still remains unclear. In order to find an appropriate combination and reduce the impact of speech distortion caused by SE, in this paper we therefore propose a joint  pre-training approach for the SE module and the self-supervised model. First, in the pre-training phase the original noisy waveform or the waveform obtained by SE is fed into the self-supervised model to learn the contextual representation, where the quantified clean speech acts as the target. Second, we propose a dual-attention fusion method to fuse the features of noisy and enhanced speeches, which can compensate the information loss caused by separately using individual modules. Due to the flexible exploitation of clean/noisy/enhanced branches, the proposed method turns out to be a generalization of some existing noise-robust ASR models, e.g., enhanced wav2vec2.0.  Finally, experimental results on both synthetic and real noisy datasets show that the proposed joint training approach can improve the ASR performance under various noisy settings, leading to a stronger noise robustness.
\end{abstract}


\section{Introduction}
\IEEEPARstart{T}{he} field of automatic speech recognition (ASR) has grown rapidly in the past few years owing to the advance in neural network-based acoustic models~\cite{6296526,graves2014towards,chorowski2015attention} and large-scale training. Compared to conventional GMM-HMM counterparts, these neural network-based end-to-end methods~\cite{8461870,kim2017joint,8462506} have a simple training procedure and can bring greater performance improvements. Although extensive progress has been achieved in ASR on clean speech, the performance would degrade sharply in the presence of background noise or under low signal-to-noise-ratio (SNR) conditions. Therefore, improving the noise robustness of ASR systems under such conditions still remains a challenge.

Regarding the noise robustness of the monaural ASR, many algorithms have been proposed recently, which can be categorized into two classes.
The first category follows the structure that combines a speech enhancement (SE) module as a front-end ASR module, in which the SE module and the ASR module can be trained separately or jointly.
For the SE module, one can use traditional methods~\cite{1163209,543199} or neural network-based methods, which can be implemented in the time domain~\cite{8462116,8683634,8707065,8701652,defossez2020real} or in the frequency domain~\cite{weninger2015speech,9103053,8462068}.
The time-domain SE reconstructs the target speech directly using the raw waveform, while the frequency-domain SE usually first estimates a mask matrix using the spectrum and then computes the spectrum of the target speech via multiplying the mask matrix by the noisy speech spectrum. In case the SE and ASR modules are trained separately~\cite{fujimoto2019one}, since the training target of SE is usually related to the instrumental speech quality, e.g., mean-square error (MSE), speech distortion, SNR, the enhanced speech might  be distorted in terms of intelligibility and thus not be optimal for ASR in terms of, e.g., word error rate (WER)~\cite{defossez2020real,7067387}. The resulting ASR performance is highly dependent on the SE module.
To alleviate the mismatch between SE and ASR modules, a method of jointly training SE and ASR modules was proposed in~\cite{7403942,fan2020gated,hu2021interactive,liu2019jointly}. Specifically,  a joint adversarial augmentation training approach was proposed in~\cite{liu2019jointly}, where a joint training framework is designed to optimize mask-based augmentation networks and attention-based encoder-decoder speech recognition networks. However, this method only uses enhanced features as input for ASR, which still suffers from the problem of speech distortion. The impact of speech distortion on ASR was analyzed in~\cite{iwamoto2022bad}, and it was shown in~\cite{fujimoto2019one,iwamoto2022bad,fan2020gated,hu2021interactive} that fusing noisy features and enhanced features can further alleviate the speech distortion issue and improve the ASR performance.

The second noise-robust ASR category mainly focuses on self-supervised pre-training. As self-supervised pre-training has shown a superiority for neural network models to leverage a large amount of unlabeled data that are widely available, many self-supervised methods for speech representation learning have been proposed recently in the speech community.
For example, autoregressive predictive coding (APC) \cite{Chung2019} was proposed to reconstruct the future frames based on the past frames.
Contrastive predictive coding (CPC) \cite{oord2018representation} and wav2vec \cite{Schneider2019} perform the next-step prediction similarly but using a contrastive loss.
Meanwhile, the contextual speech representations can be learned from the unlabeled speech data by reconstructing the masked input speech frames~\cite{liu2021tera,ling2020deep}.
For contextual speech representation learning, the bidirectional Transformer (BERT) structure~\cite{vaswani2017attention}  is utilized in~\cite{liu2021tera},  and the bidirectional long short-term memory (LSTM) structure is used in~\cite{ling2020deep}.
Vq-wav2vec \cite{baevski2019vq} utilizes a quantization module to extract discrete semantic units from unlabeled speech data and then uses BERT to perform contextual modeling on the extracted units.
Wa2vec2.0 \cite{NEURIPS2020_92d1e1eb} employs a convolutional neural network (CNN) to extract local features from the raw waveform, which are then input to the BERT module to perform mask prediction using a contrastive loss.
HUBERT~\cite{9585401} performs offline clustering on representations, which enables a direct prediction of the clustering label of the masked positions. In addition, on the basis of the HUBERT framework, UniSpeech-SAT~\cite{chen2021unispeech} and WavLM~\cite{chen2021wavlm} were proposed  to boost the spoken content information and speaker identity using an utterance-wise contrastive loss and an utterance mixing data augmentation method, respectively.

It was shown in \cite{kawakami2020learning,riviere2020unsupervised} that a modified CPC pre-trained model can be transferred well across domains, and larger pre-training datasets lead the ASR model to be much more robust against the domain shift.
The robust wav2vec2.0 proposed in \cite{hsu2021robust} reveals the impact of domain mismatch on the self-supervised speech representation learning.
The problem-agnostic speech encoder (PASE+) that was proposed in \cite{ravanelli2020multi} introduces online speech data augmentation modules for self-supervised learning and obtains a good performance in noisy environments. Wav2vec-switch~\cite{wang2021wav2vec} encodes the noise robustness into contextualized representations of speech via contrastive learning. In~\cite{wang2021improving}, an extra reconstruction module was used for auxiliary learning to improve the noise robustness of the learned representation.  Consistency contrastive learning method~\cite{gao2021data} was utilized for acoustic pre-training, which shows a WER improvement on both the in-domain data and the out-domain data. By taking the quantized clean speech as the training target for pre-training, we proposed an enhanced wav2vec2.0 in~\cite{zhu2022noise}, resulting in a better speech representation.

In principal, both SE and self-supervised pre-training based methods can improve the noise robustness of ASR to some extent, but from different perspectives.
The SE based methods require careful adjustments for network training, while self-supervised methods can utilize more unlabeled noisy data to pre-train and then directly fine-tune the pre-trained models to improve the noise robustness, which usually have less speech distortion issues compared to the former.
A new question arises  {\it whether an appropriate combination can improve the noise robustness consistently and reduce the impact of speech distortion caused by SE}.
To our knowledge, there are few works  on this question, e.g.,
in~\cite{chang2022end}, a trained SE module, a trained self-supervised module for feature extraction, and a trained ASR module are integrated into an end-to-end framework, which is then fine-tuned altogether and achieves the best performance on the CHiME-4~\cite{vincent20164th} single-channel ASR task.
However, in case this model is trained using random initialization, the performance or convergence cannot be guaranteed due to the network depth and gradient back propagation.

In this work, we therefore explore a joint pre-training method incorporating  SE  and self-supervised training to improve the noise robustness of ASR.
This work is an extension of the conference paper in~\cite{zhu2022noise}, where only the structure of the enhanced wav2vec2.0 model and  results on the synthetic data were shown.
Compared to~\cite{zhu2022noise}, the contribution of this paper is threefold.
First, we propose a joint pre-training approach for the DEMUCS~\cite{defossez2020real} based time-domain SE module and the enhanced wav2vec2.0 based self-supervised model. In the pre-training stage, the original noisy waveform or the waveform after SE is fed to the self-supervised model to learn the contextual representation, and the quantized clean speech provides the target for the pre-training model. Second, we propose a dual-attention fusion module to fuse the features of the noisy and enhanced speeches, such that  the information loss caused by the SE can be compensated.
Third, the proposed pre-training model is validated using both synthetic data and the real noisy CHiME-4 dataset.
Results show that adding consistency constraints on the features output from the feature encoder and using the quantized clean features as targets for the pre-training model are beneficial for  the noise robustness of the ASR model on both datasets. In addition, it is shown that the SE module and the self-supervised model can jointly improve the ASR performance in noisy scenes, and the dual-attention fusion module can reduce the effect of distortion caused by SE.
Finally, we find that using a pre-trained model for initialization can also mitigate the speech distortion and improve the ASR performance, which is theoretically analyzed from the viewpoint of numerical optimization.

The remainder of the paper is arranged as follows.
In Section~\uppercase\expandafter{\romannumeral2}, we present the proposed noise-robust self-supervised pre-training model including the enhanced wav2vec2.0 model, the SE module, dual-attention fusion and the joint pre-training technique. Experimental setups are described in Section~\uppercase\expandafter{\romannumeral3}, followed by extensive experimental results and analysis  in Section~\uppercase\expandafter{\romannumeral4}. Finally,  Section~\uppercase\expandafter{\romannumeral5} concludes this work.

\section{Methodology}

In this section, we will present the proposed self-supervised pre-training model, including the enhanced wav2vec2.0 (EW2), the time-domain SE module, the proposed dual-attention fusion module and the joint pre-training technique.
\subsection{The enhanced wav2vec2.0 (EW2) model}

\begin{figure}[!t]
  \centering
  \includegraphics[width=0.45\textwidth]{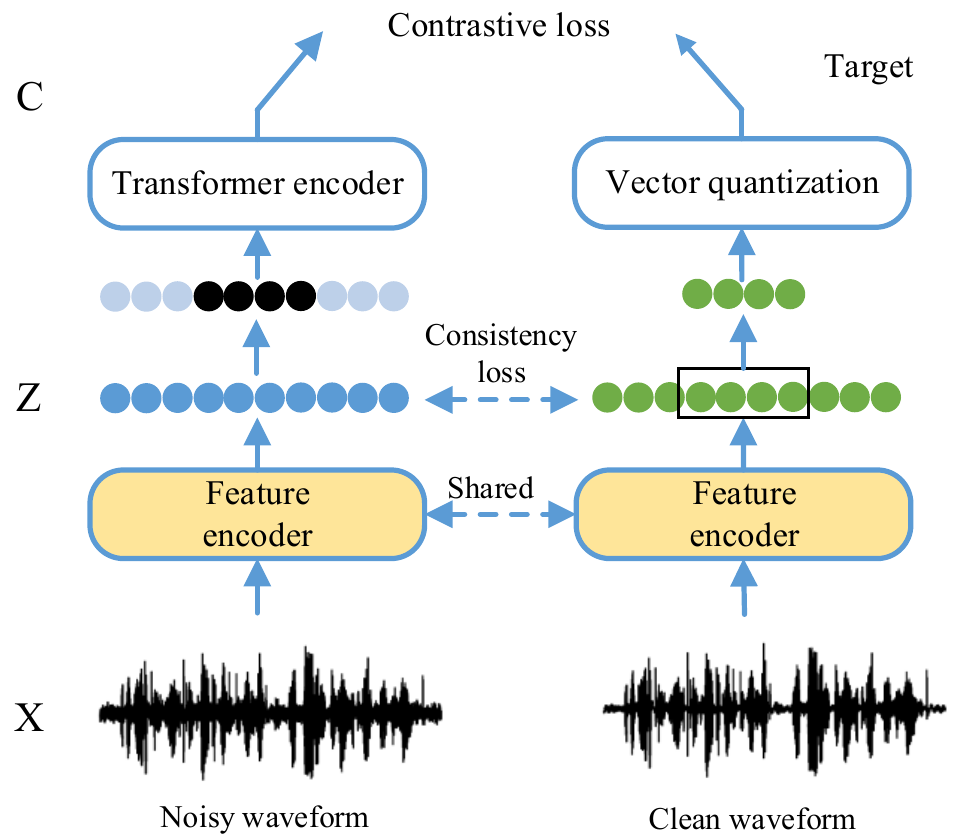}
  \caption{The structure of the enhanced wav2vec2.0 model.}
  \label{fig:enhancedwav2vec2}
\end{figure}

The proposed enhanced wav2vec2.0 model is based on the classic wav2vec2.0~\cite{NEURIPS2020_92d1e1eb}, and the corresponding model structure is shown in Fig.~\ref{fig:enhancedwav2vec2}, which also consists of a feature encoder $f:X \mapsto Z$, a Transformer encoder $g: Z \mapsto C $ and a vector quantization (VQ) $g: Z \mapsto Q$.
The feature encoder utilizes seven CNN layers and the Transformer encoder contains twelve transformer blocks.
Specifically,  the shared feature encoder extracts the noisy features $Z_{\rm noisy}$ from the raw noisy waveform $X_{\rm noisy}$ and extracts the clean features $Z_{\rm clean}$ from the raw clean waveform $X_{\rm clean}$, respectively, which can then be described as
\begin{equation}
  Z_{\rm noisy} = f(X_{\rm noisy}),\quad
  Z_{\rm clean} = f(X_{\rm clean}).
  \label{eq1}
\end{equation}
We mask a certain proportion of the noisy features $Z_{\rm noisy}$ by replacing with a learnable vector at the masked position.
Then, the high-level noisy contextualized representations $C_{\rm noisy}$ are learned from the noisy features $Z_{\rm noisy}$ by the Transformer encoder, which is given by
\begin{equation}
  C_{\rm noisy} = g(Z_{\rm noisy}).
  \label{eq3}
\end{equation}
The corresponding clean features $Z_{\rm clean}$ are discretized into $q_{\rm clean}$ via a VQ module, which are then used as clean targets in the contrastive objective, i.e.,
\begin{equation}
  q_{\rm clean} = VQ(Z_{\rm clean}).
  \label{eq4}
\end{equation}
The motivation of using clean features as the target originates from the expectation that the model can learn clean speech representations from noisy features.
The involved  VQ module is implemented using the product quantization \cite{jegou2010product}.
Specifically, the VQ module first maps the clean features $Z_{\rm clean}$ to logits $\mathbf{l} \in \mathbb{R}^{G \times V} $, where $G$ represents the number of codebooks and $V$ the number of entries in each codebook.
The gumbel softmax function \cite{jang2016categorical} is then used to select discrete codebook entries in a fully differentiable way. As a result, for a given frame $Z_{{\rm clean}_{t}}$ at time $t$ , we can select one entry from each codebook, concatenate the resulting vectors $e_1,...,e_G$ and apply a linear transformation to obtain $ q_{{\rm clean}_{t}} $. The loss function can therefore be formulated as
\begin{equation}
  L = L_{m}+\alpha L_{d} + \beta L_{f} + \gamma L_{c},
  \label{eq5}
\end{equation}
where
\begin{align}
  L_{m} &= -\log \frac{\exp({\rm sim}(C_{{\rm noisy}_t},q_{{\rm clean}_t})/\kappa)}{\sum_{\tilde{q} {\sim} Q_{t}}\exp({\rm sim}(C_{{\rm noisy}_t},\tilde{q})/\kappa)},
  \label{eq6} \\
  L_{d} &= \frac{1}{GV}\sum_{g=1}^{G}\sum_{v=1}^{V}\overline{p}_{g,v}\log \overline{p}_{g,v},
  \label{eq7}\\
  \overline{p}_{g,v}&=\frac{\exp(\overline{l}_{g,v}+n_{v})/\tau}{\sum_{k=1}^{V}\exp(\overline{l}_{g,k}+n_{k})/\tau},
  \label{eq8} \\
  L_{c} &= \left\| Z_{{\rm noisy}_{t}}-Z_{{\rm clean}_{t}} \right\|_2,
  \label{eq9}
\end{align}
which applies to any time index $t$.
It is clear that the total loss function is the weighted summation over four terms depending on the parameters $\alpha$, $\beta$ and $\gamma$. In (\ref{eq5}), $L_{m}$ is the contrastive loss, which enables the model to distinguish between the true quantized clean features $q_{{\rm clean}_t}$ and a set of $K+1$ quantized candidate features $\tilde{q} \in Q_{t}$. The quantized candidate features $\tilde{q}$ contains $q_{{\rm clean}_t}$ and $K$ distractors. The diversity loss $L_{d}$ aims to increase the use of quantized codebook features, and $L_{f}$ is an $\ell_2$ penalty over the outputs of the feature encoder.  In (\ref{eq6}), sim stands for the cosine similarity between two vectors and $\kappa$ is a temperature. In (\ref{eq7}), $\overline{p}_{g,v}$ represents the probability of choosing the $v$-th codebook entry for group $g$ across a batch of utterances, where $\tau$ is a temperature. In (\ref{eq8}), $\overline{l}_{g,v}$ stands for the average logits $\mathbf{l}$ across utterances in a batch. In order to ensure the consistency between the clean features and the noisy features corrupted by noise, we additionally introduce a consistency loss $L_{c}$  in (\ref{eq9}), which measures the Euclidean distance between noisy features $Z_{\rm noisy}$ and clean features $Z_{\rm clean}$.

\subsection{Speech enhancement (SE) module}

Many time-domain and frequency-domain SE models have been proposed in literature.
The superiority of time-domain SE methods is that no specific operations on the phase information are required for generating estimated signals and the combination with existing self-supervised models is straightforward. In principal, any type of SE models can be employed in the proposed self-supervised pre-training approach. Without loss of generality, we choose the time-domain DEMUCS model~\cite{defossez2020real} for SE, which is built by optimizing both the time-domain and frequency-domain loss functions.
DEMUCS consists of a five-layer convolutional encoder, a two-layer LSTM network and a five-layer convolutional decoder. The encoder includes several convolution layers (where the number of input channels in the first convolution layer is $H_{\rm se}$, the convolution kernel size of the $i$-th encoder layer is $K_{\rm se}$, the convolution stride is $S_{\rm se}$ and the number of output channels is $2^{i-1} H_{\rm se}$), a ReLU activation and a 1$\times$1 convolution with $2^{i} H_{\rm se}$ output channels.
Then, a gated linear units (GLU) activation converts the output channels back to $2^{i-1} H_{\rm se}$.
Next, the LSTM network is applied to the output of the encoder to model the sequential hidden state.
For the decoder, the number of input channels of the i-th transposed convolution layer is $2^{i-1} H_{\rm se}$, followed by a 1$\times$1 convolution with $2^{i} H_{\rm se}$ channels and a GLU activation that outputs $2^{i-1} H_{\rm se}$ channels.
It is worth noting that  a skip connection between the outputs of the $i$-th encoder layer and the $i$-th decoder layer is required.

The DMEUCS considers both time-domain and frequency-domain loss functions, where the time-domain loss function on the waveform is $\ell_1$ loss and the frequency-domain one relies on the multi-resolution short-time Fourier transform (STFT) coefficients.
Suppose $X$ and $X_{\rm en}$ represent the clean speech waveform signal and the enhanced speech waveform signal, respectively, the loss function can be written as
\begin{equation}
  L_{\rm SE} = \frac{1}{T} \left(\left\| X - X_{\rm en}  \right\|_1 + \sum_{i=1}^{M}L_{\rm stft}^{(i)}(X, X_{\rm en})\right),
  \label{eq10}
\end{equation}
where $M$ is the number of STFT losses, the first term represents the time-domain loss on the waveform, and the second term denotes the loss on STFT coefficients, which is defined as the sum of spectral convergence (sc) loss and magnitude loss, i.e.,
\begin{equation}
  L_{\rm stft}(X,X_{\rm en}) = L_{\rm sc}(X,X_{\rm en}) + L_{\rm mag}(X,X_{\rm en}),
  \label{eq11}
\end{equation}
where
\begin{align}
  L_{\rm sc}(X,X_{\rm en}) &= \frac{\left\| | {\rm STFT}(X)| - | {\rm STFT}(X_{\rm en})| \right\|_F}{\left\|  {\rm STFT}(X) \right\|_F},
  \label{eq12}\\
  L_{\rm mag}(X,X_{\rm en}) &= \frac{1}{T} \left\| \log |{\rm STFT}(X)|-\log |{\rm STFT}(X_{\rm en})|  \right\|_1,
  \label{eq13}
\end{align}
where $\left\| \cdot \right\|_F$ and $\left\| \cdot \right\|_1$ denote  the Frobenius  and $\ell_1$ norms, respectively, and ${\rm STFT}(\cdot)$ returns the STFT coefficient matrix.
As such, the multi-resolution $L_{\rm stft}^{(i)}$ can utilize the STFT loss at different resolutions with the number of FFT bins, e.g., $\in$ \{512, 1024, 2048\}, hop sizes $\in$ \{50, 120, 240\}, and window lengths $\in$ \{240, 600, 1200\}. For more details about DEMUCS, please refer to~\cite{defossez2020real}.

\subsection{The dual-attention fusion module}

\begin{figure}[!t]
  \centering
  \includegraphics[width=0.45\textwidth]{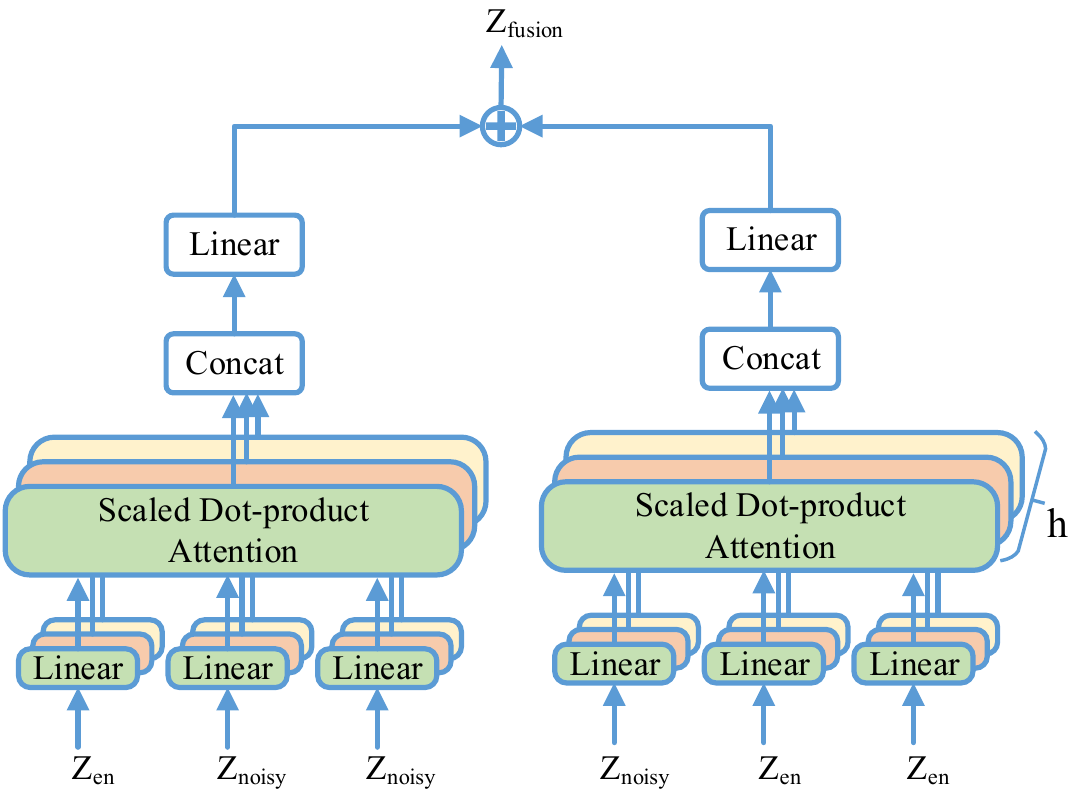}
  \caption{An illustration of the proposed dual-attention fusion module. }
  \label{fig:cfm}
\end{figure}

In order to learn the complementary information of the noisy features and enhanced features, we propose a dual attention fusion module on the basis of the classic attention mechanism~\cite{vaswani2017attention}, which is shown in Fig.~\ref{fig:cfm}.
In the left branch, the enhanced features $Z_{\rm en}$ and the noisy features $Z_{\rm noisy}$ are used to calculate the attention weight, and the output is the weighted sum of the noisy features $Z_{\rm noisy}$.
Similarly, in the right branch, attention weights are calculated from noisy features $Z_{\rm noisy}$ and enhanced features $Z_{\rm en}$, and the output is the weighted sum of enhanced features $Z_{\rm en}$. The fused feature is then the summation over the outputs of the two branches,  given by
\begin{equation}
\begin{aligned}[b]
Z_{\rm fusion} = &\text{Linear}(\text{Multihead}(Z_{\rm en},Z_{\rm noisy},Z_{\rm noisy}))+\\
&\text{Linear}(\text{Multihead}(Z_{\rm noisy},Z_{\rm en},Z_{\rm en})),
\label{eq14}
\end{aligned}
\end{equation}
which involves a linear mapping layer, and the multi-head attention is formulated as
\begin{equation}
\text{Multihead}(Z_{Q},Z_{K},Z_{V})=\text{Concat}(h_{1},...,h_{n})W^O,
\label{eq15}
\end{equation}
where $Z_{Q}$, $Z_{K}$ and $Z_{V}$ can be replaced by $Z_{\rm noisy}$ or $Z_{\rm en}$. The scaled dot-product attention is calculated as
\begin{equation}
\begin{aligned}[b]
h_{i} &= \text{Attention}(Z_{Q}W_{i}^Q,Z_{K}W_{i}^K,Z_{V}W_{i}^V)\\
&=\text{softmax}\left(\frac{Z_{Q}W_{i}^Q({Z_{K}W_{i}^K})^T}{\sqrt{d_{k}}}\right)Z_{V}W_{i}^V
\label{eq16}
\end{aligned}
\end{equation}
where $d_{k}$ equals the dimension $d_z$ of $Z_{\rm noisy}$ over $h$ (i.e., $d_k=d_z/h$), and $W_{i}^Q \in \mathbb{R}^{d_{z} \times d_{k}}$, $W_{i}^K \in \mathbb{R}^{d_{z} \times d_{k}}$, $W_{i}^V \in \mathbb{R}^{d_{z} \times d_{k}}$ and $W_{i}^O \in \mathbb{R}^{d_{z} \times d_{z}}$ are learnable parameters.

\subsection{Joint training for SE and EW2}

\begin{figure*}[!ht]
  \centering
  \includegraphics[width=0.8\textwidth]{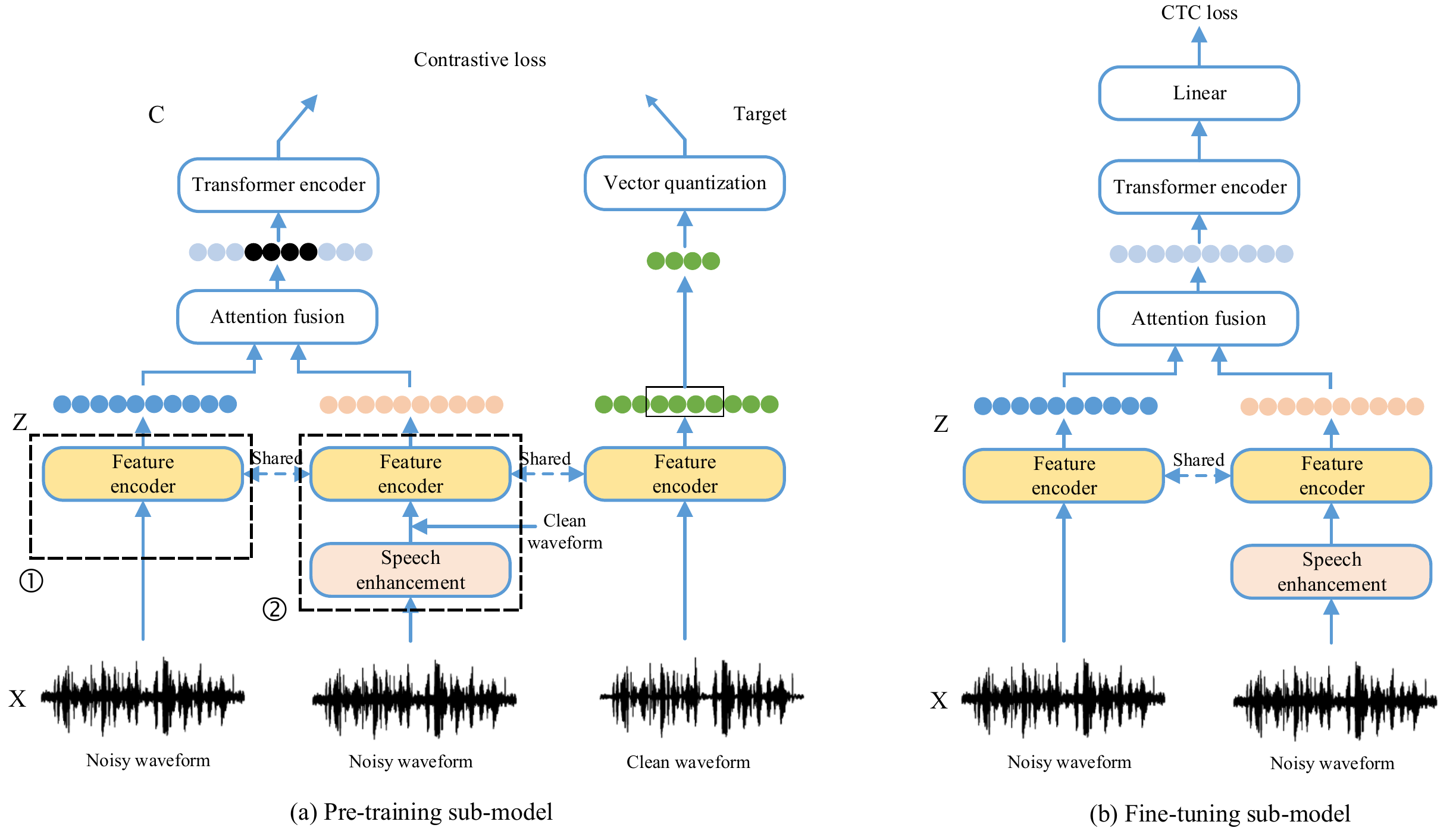}
  \vspace{-0.1cm}
  \caption{An illustration of the proposed joint pre-training model: (a) pre-training sub-model and (b) fine-tuning sub-model.}
  \label{fig:robustse}
\end{figure*}

The model structure of the proposed pre-training approach is shown in Fig.~\ref{fig:robustse}(a), where both the SE  module in Section II-B and the back-end self-supervised module in Section II-A are included.
As shown in (\ref{eq1}), the input raw noisy waveform $X_{\rm noisy}$ and raw clean waveform $X_{\rm clean}$ are sent to the shared feature encoder to output the noisy features $Z_{\rm noisy}$ and the clean features $Z_{\rm clean}$, respectively. The noisy speech $X_{\rm noisy}$ is also sent to the SE module to obtain the enhanced speech $X_{\rm en}$, i.e.,
\begin{equation}
  X_{\rm en} = \text{enhancer}(X_{\rm noisy}),
  \label{eq17}
\end{equation}
 which is then sent to the shared feature encoder to construct the enhanced features $Z_{\rm en}$ as
\begin{equation}
  Z_{\rm en} = f(X_{\rm en}).
  \label{eq18}
\end{equation}

The noisy features and enhanced features are fused using a convolutional fusion module to obtain the final fused features as in (\ref{eq14}). The fused feature $Z_{\rm fusion}$ is randomly masked at a certain proportion and then sent to the Transformer module to learn the contextual representation $C_{\rm fusion}$, which is given by
\begin{equation}
  C_{\rm fusion} = g(Z_{\rm fusion}).
  \label{eq19}
\end{equation}
On the other hand, the corresponding clean features $Z_{\rm clean}$ are discretized into $q_{\rm clean}$ via a VQ module following $Z \mapsto Q $ in (\ref{eq4}), which are then used as clean targets in the contrastive objective. It is worth mentioning that the clean speech signal is only used in the pre-training stage but not in the fine-tuning stage.
The overall loss function of the proposed joint training strategy can thus be formulated as
\begin{equation}
  L_{\rm total} = L_{\rm contrastive} + \xi L_{\rm SE}.
  \label{eq20}
\end{equation}
where $L_{\rm SE}$ was given in (\ref{eq10}) and $L_{\rm contrastive}$ can be formulated similarly as that in Section II-A, e.g.,
\begin{equation}
  L_{\rm contrastive} = L_{ms}+\alpha L_{d} + \beta L_{f} + \gamma L_{cs},
  \label{eq21}
\end{equation}
where
\begin{align}
  L_{ms} &= -\log \frac{\exp({\rm sim}(C_{{\rm fusion}_t},q_{{\rm clean}_t})/\kappa)}{\sum_{\tilde{q} {\sim} Q_t}\exp({\rm sim}(C_{{\rm fusion}_t},\tilde{q})/\kappa)},
  \label{eq22} \\
  L_{cs} &= \left\| Z_{{\rm noisy}_t}-Z_{{\rm clean}_t} \right\|_2 + \left\| Z_{{\rm en}_t}-Z_{{\rm clean}_t} \right\|_2.
  \label{eq23}
\end{align}
In (\ref{eq22}), $L_{ms}$ is the contrastive loss between the contextual representation $C_{\rm fusion}$ and a set of quantized candidate features $\tilde{q}$. In (\ref{eq23}), $L_{cs}$ is the consistency loss function between $Z_{\rm noisy}$, $Z_{\rm en}$ and $Z_{\rm clean}$. The loss function of $L_{d}$ and $L_{f}$ keep the same as in (\ref{eq5}), and $\alpha$, $\beta$, $\gamma$, $\xi$ in (\ref{eq20}) and (\ref{eq21}) are the weighting parameters, which can be set empirically.

It should be noted that based on the pre-training structure in Fig.~\ref{fig:robustse}(a), apart from the proposed joint pre-training framework (i.e., all three branches and the dual-attention fusion are involved, thus termed by EW2+SEW2), one can further construct several noise-robust ASR models. For example, in case only the left branch \ding{172} and the right clean speech branch are used, the proposed model reduces to the off-the-shelf EW2 model~\cite{zhu2022noise}.
In case the left branch \ding{173} and the right clean speech branch are available, the proposed model is equivalent to the SE-based EW2 (SEW2). In case the dual-attention fusion module is replaced by simple concatenation, the model reduces to EW2+SEW2 (concat), which would have less parameters than the proposed EW2+SEW2. Therefore, the proposed method can be seen as a generalization of EW2 and SEW2 that make use of the clean speech for noise-robust ASR. In Section~\uppercase\expandafter{\romannumeral4}, these variants will also be compared.

In the fine-tuning stage, we removed the right clean speech branch in Fig.~\ref{fig:robustse}(a), and the corresponding model structure is shown in Fig.~\ref{fig:robustse}(b).
We  add an additional linear mapping layer to the output of the Transformer encoder and optimize the whole model using the connectionist temporal classification (CTC)~\cite{graves2006connectionist} loss function with a small amount of labeled data.
Note that in case the noisy and clean branches are used in the pre-training stage (i.e., EW2), the same noisy branch is required for fine-tuning. This applies to other variants, that is, the branch combination for pre-training and fine-tuning should be kept consistent.

\section{Experimental setup}
In this section, the datasets for model validation and experimental configuration on pre-training and fine-tuning will be described in  detail.
\subsection{Dataset description}
\textbf{LibriSpeech:} In order to facilitate a fair comparison with existing approaches, the data usage throughout experiments keeps exactly the same as that in \cite{9414027}. Specifically, we utilize the LibriSpeech \cite{7178964} train-clean-100 subset and train-clean-360 subset as the clean speech training set and standard dev-clean subset as the validation set. For the model training and validation, we randomly select noise samples and mix with the clean speech additively at an arbitrarily chosen SNR from [0 dB, 25 dB] to generate noisy data.
The noisy test set is downloaded from the website\footnote{https://github.com/archiki/Robust-E2E-ASR}, which includes 4200 noisy audio streams generated by selecting 120 clean speech samples from the test-clean subset of LibriSpeech and then mixing with noises at different SNRs $\in$ \{0, 5, 10, 15, 20\} dB.
The noise data used in experiments originate from FreeSound \cite{font2013freesound}, which are sampled at a sampling frequency of 16 kHz. The noise type is divided into two categories, i.e., A and B. The type-A noise is relatively stationary, including `Car', `Metro' and `Traffic' noise, and the type-B noise is relatively non-stationary, including `Babble', `Airport/Station', `Cafe' and `AC/Vacuum' noise. Each noise type has 10 and 8 different audio streams in the training and test sets, respectively. The length of the noise dataset is around 2 hours in total.
In addition, we also utilize the NoiseX-92~\cite{varga1993assessment} noise dataset at the pre-training stage for supplementary experiments.

 \textbf{CHiME-4:} To further verify the effectiveness of the proposed method on real noisy data, we conduct experiments on the CHiME-4 challenge\footnote{http://spandh.dcs.shef.ac.uk/chime\_challenge/CHiME4/index.html} dataset \cite{vincent20164th}. The CHiME-4 dataset is related to the text from the Wall Street Journal (WSJ0) corpus, which was collected using a six-channel distant microphone array and a close-talk microphone for data recording when the volunteer is reading the corresponding text.  This dataset contains two types of noisy speech, i.e., real and artificially simulated noisy speech. The real noisy data was recorded in the noisy environments including bus, cafe, pedestrian area, and street junction, and the artificial simulation data was synthesized by mixing the four kinds of noise with the clean speech. The training set contains 1600 real recorded and 7138 simulated noisy utterances, the validation set contains 1640 real recorded and 1640 simulated noisy utterances, and the test set contains 1320 real recorded and 1320 simulated noisy utterances. All these three data subsets were constructed from four different speakers.
 As the focus of this paper is  on the single-channel ASR,  the data of all six channels will be used as training data in the fine-tuning stage, and one-channel track real noisy data is used as validation and test sets, which is similar to the data configuration in \cite{wang2021wav2vec,wang2021improving}.

\begin{table*}[]
\caption{ The performance comparison of different methods on type-A and type-B noise test sets at different SNRs, where ``No" means that the pre-training step is not included and wav2vec2.0 is fine-tuned directly from scratch with the target dataset. The pre-training and fine-tuning can be performed on different datasets.}
\label{tab:enhtable1}
\centering
\begin{tabular}{l|c|c|ccccccccc}
\hline
\multirow{4}{*}{\textbf{Method}}   & \multirow{4}{*}{\textbf{Pre-train}} & \multirow{4}{*}{\textbf{Fine-tune}} & \multicolumn{9}{c}{\textbf{WER under SNR (dB)}}                                                                                                                                                                                                                                                                                                                                                                    \\ \cline{4-12}
                          &                            &                            & \multicolumn{4}{c|}{\textbf{Type-B noise}}                                                                                                                                                                                             & \multicolumn{3}{c|}{\textbf{Type-A noise}}                                                                & \multicolumn{1}{c|}{\multirow{2}{*}{\textbf{Average}}} & \multirow{2}{*}{\textbf{Clean}} \\ \cline{4-10}
                          &                            &                            & \multicolumn{1}{c|}{\textbf{Babble}}    & \multicolumn{1}{c|}{\begin{tabular}[c]{@{}c@{}}\textbf{Airport/}\\ \textbf{Station}\end{tabular}} & \multicolumn{1}{c|}{\begin{tabular}[c]{@{}c@{}}\textbf{AC/}\\ \textbf{Vacuum}\end{tabular}} & \multicolumn{1}{c|}{\textbf{Cafe}}      & \multicolumn{1}{c|}{\textbf{Traffic}}   & \multicolumn{1}{c|}{\textbf{Metro}}     & \multicolumn{1}{c|}{\textbf{Car}}       & \multicolumn{1}{c|}{}                         &                        \\ \cline{4-12}
                          &                            &                            & \multicolumn{1}{c|}{\textbf{0$\sim$20}} & \multicolumn{1}{c|}{\textbf{0$\sim$20}}                                                  & \multicolumn{1}{c|}{\textbf{0$\sim$20}}                                            & \multicolumn{1}{c|}{\textbf{0$\sim$20}} & \multicolumn{1}{c|}{\textbf{0$\sim$20}} & \multicolumn{1}{c|}{\textbf{0$\sim$20}} & \multicolumn{1}{c|}{\textbf{0$\sim$20}} & \multicolumn{1}{c|}{\textbf{0$\sim$20}}                & -                      \\ \hline
Baseline~\cite{9414027}                  & No                         & Clean                      & 87.70                          & 72.06                                                                           & 70.58                                                                     & \multicolumn{1}{c|}{58.74}     & 51.94                          & 46.34                          & \multicolumn{1}{c|}{25.16}     & 58.93                                         & \textbf{10.3}                   \\
DEMUCS~\cite{9414027}                    & FreeSound                  & FreeSound                  & 45.56                          & 36.98                                                                           & 38.20                                                                     & \multicolumn{1}{c|}{27.02}     & 26.46                          & 23.22                          & \multicolumn{1}{c|}{16.02}     & 30.49                                         & 10.9                   \\
AvT~\cite{9414027}                       & No                         & FreeSound                  & 43.42                          & 35.32                                                                           & 36.62                                                                     & \multicolumn{1}{c|}{27.06}     & 27.88                          & 24.28                          & \multicolumn{1}{c|}{17.76}     & 30.33                                         & 13.1                   \\ \hline
\multirow{5}{*}{Wav2vec2.0~\cite{zhu2022noise}} & No                         & Clean                      & 81.74                          & 70.20                                                                           & 67.88                                                                     & \multicolumn{1}{c|}{57.40}     & 50.24                          & 46.58                          & \multicolumn{1}{c|}{23.60}     & 56.81                                         & 11.0                   \\
                          & No                         & FreeSound                  & 59.40                          & 50.88                                                                           & 49.96                                                                     & \multicolumn{1}{c|}{43.58}     & 40.36                          & 37.78                          & \multicolumn{1}{c|}{29.94}     & 44.56                                         & 25.0                   \\
                          & Clean                      & FreeSound                  & 47.50                          & 39.68                                                                           & 38.84                                                                     & \multicolumn{1}{c|}{31.14}     & 29.22                          & 27.44                          & \multicolumn{1}{c|}{18.24}     & 33.15                                         & 14.0                   \\
                          & FreeSound                  & FreeSound                  & 39.56                          & 32.50                                                                           & 34.94                                                                     & \multicolumn{1}{c|}{25.22}     & 24.52                          & 22.48                          & \multicolumn{1}{c|}{16.24}     & 27.92                                         & 13.5                   \\
                          & NoiseX-92                  & FreeSound                  & 43.18                          & 34.16                                                                           & 38.80                                                                     & \multicolumn{1}{c|}{27.06}     & 25.58                          & 24.64                          & \multicolumn{1}{c|}{18.08}     & 30.21                                         & 16.1                   \\ \hline
\multirow{2}{*}{EW2~\cite{zhu2022noise}}      & FreeSound                  & FreeSound                  & 33.88                          & 27.36                                                                           & 27.94                                                                     & \multicolumn{1}{c|}{22.08}     & 20.94                          & 19.84                          & \multicolumn{1}{c|}{14.88}     & 23.85                                         & 12.3                   \\
                          & NoiseX-92                  & FreeSound                  & 38.80                          & 30.12                                                                           & 30.94                                                                     & \multicolumn{1}{c|}{22.72}     & 22.06                          & 20.96                          & \multicolumn{1}{c|}{15.98}     & 25.94                                         & 14.3                   \\ \hline
SEW2                      & No                         & FreeSound                  & 50.50                          & 42.68                                                                           & 41.96                                                                     & \multicolumn{1}{c|}{34.88}     & 32.62                          & 31.60                          & \multicolumn{1}{c|}{22.86}     & 36.73                                         & 19.1                   \\
SEW2                      & FreeSound                  & FreeSound                  & 39.64                          & 33.90                                                                           & 33.28                                                                     & \multicolumn{1}{c|}{26.90}     & 25.54                          & 24.88                          & \multicolumn{1}{c|}{18.26}     & 28.91                                         & 15.4                   \\
EW2 + SEW2 (concat)       & No                         & FreeSound                  & 47.44                          & 39.64                                                                           & 41.60                                                                     & \multicolumn{1}{c|}{32.04}     & 31.82                          & 30.24                          & \multicolumn{1}{c|}{22.32}     & 35.01                                         & 18.5                   \\
EW2 + SEW2                & No                         & FreeSound                  & 44.54                          & 36.00                                                                           & 36.12                                                                     & \multicolumn{1}{c|}{29.28}     & 27.00                          & 25.64                          & \multicolumn{1}{c|}{18.76}     & 31.05                                         & 15.6                   \\
EW2 + SEW2 (concat)       & FreeSound                  & FreeSound                  & 33.24                          & 26.91                                                                           & 27.52                                                                     & \multicolumn{1}{c|}{21.78}     & 20.63                          & 19.62                          & \multicolumn{1}{c|}{14.60}     & 23.47                                         & 12.2                   \\
EW2 + SEW2                & FreeSound                  & FreeSound                  & \textbf{31.55}                          & \textbf{25.90}                                                                           & \textbf{26.83}                                                                     & \multicolumn{1}{c|}{\textbf{21.22}}     & \textbf{19.75}                          & \textbf{19.04}                          & \multicolumn{1}{c|}{\textbf{14.25}}     & \textbf{22.65}                                         & 12.2                   \\ \hline
\end{tabular}
\end{table*}

\subsection{Model configuration}

\textbf{Pre-train on 100 hours unlabeled data:} This configuration means that we pre-train on 100 hours of clean-noisy paired speech data from the train-clean-100 subset of LibriSpeech, where the noisy speech data is generated by using clean speech dynamically mixing with noise. The model structure is implemented using the fairseq toolkit\footnote{https://github.com/pytorch/fairseq}. In detail, the feature encoder consists of seven convolutional layers and the channel number of the convolution module is 512. The stride and kernel sizes of the convolution module are (5, 2, 2, 2, 2, 2, 2) and (10, 3, 3, 3, 3, 2, 2), respectively. Therefore, the frame shift of the output $Z_{\rm noisy}$ of the feature encoder is 20 ms and its receptive field is 25 ms.

For the Transformer encoder module, we utilize 12  Transformer encoder layers and each contains a self-attention module and a feed forward module. The dimension of the self-attention module is 512, and 8 heads are utilized. The dimension of the feed forward module is 512, and the inner dimension is 2048. For the VQ module, we set $G$ = 2 and $V$ = 320, and the dimension of each entry is 128. The model size including all
parameters is around 45 M. For masking, we sample at all time steps at a probability of $p$ = 0.065 and mask the subsequent $M$ = 10 time steps. For the loss function, the temperature $\kappa$ is set to be 0.1, and $\tau$ is annealed from 2 to 0.5 with a coefficient of 0.999995 in terms of iterations. The parameters $\alpha$, $\beta$, $\gamma$ and $\xi$ are set to be  0.1, 10, 1 and 0.1, respectively. The number of distractors $K$ equals 100.

\textbf{Pre-train on 460 hours unlabeled data:} This configuration means that we pre-train on 460 hours of clean-noisy paired speech from the  train-clean-100 subset and train-clean-360 subset of LibriSpeech, where the noisy speech data is generated similarly as before. The main difference from the previous configuration lies in the size of the model parameters, where the size of the model parameters is around 95 M. The dimension of the self-attention module is 768, and the number of attention heads is 12. The dimension of the feed forward module is 768, and the inner dimension is 2048. The rest parameters keep the same as the first configuration.

\textbf{Fine-tune on 100 hours labeled data:}
During fine-tuning, we use the noisy speech at different SNRs to fine-tune the model, where the generation of noisy data remains the same as that in the pre-training phase. The model also uses a data augmentation method similarly to SpecAugment~\cite{park2019specaugment}, where a time-frequency mask is applied to the feature output by the feature encoder. The time masking probability is 0.065, and 10 consecutive frames are masked. The frequency masking probability is 0.05, and 32 consecutive channels are masked. The modeling unit has 30 characters, including 26 letters and 4 special symbols.
After fine-tuning, we decode on clean test sets and noisy test sets without any language model (LM) and calculate the WER for performance evaluation.

\textbf{Pre-train and Fine-tune on CHiME-4 data:} For a fair comparison with  existing approaches, we adopt the same data configuration therein. Both real and simulated data from all channels except for the second microphone channel are utilized for pre-training and fine-tuning.
Due to the small data size of CHiME-4 data, pre-training with random initialization cannot guarantee an acceptable performance, so we continue pre-training 50k updates based on the public pre-trained model\footnote{https://dl.fbaipublicfiles.com/fairseq/wav2vec/wav2vec\_small.pt} and then fine-tune 20k updates with labeled data. Note that the public pre-trained model was pre-trained using 960 hours speech data of LibriSpeech, which has about 95 M parameters. We also apply the time-frequency masking for data augmentation and fine-tune the model with the CTC loss function. Since the results from e.g.,~\cite{wang2021wav2vec,wang2021improving} on the CHiME-4 dataset employ LMs, we train a Transformer-based word-level LM with a vocabulary of 65,000 using the text portion of the WSJ corpus.

The Transformer-based LM contains a 16-layer encoder, where both the encoder and self-attention dimensions are 512 and the inner dimension of the feed forward neural network is 2048. The loss function is cross-entropy, which is optimized using the Adam optimizer. The training process of LM is implemented using the fairseq toolkit. We utilize a simple shallow fusion to integrate external LMs, which are integrated by calculating the weighted distributions of two modeling units, i.e., one from the ASR model and the other from the external Transformer LM. The weight of the LM is 1.0, and the beam size is 500.

For the SE module, $H_{\rm se}$, $K_{\rm se}$ and $S_{\rm se}$ are set to be 64, 8 and 4, respectively. For the dual-attention fusion module, the dimension of $Z_{\rm noisy}$ is set to be 512, the number of attention head $h$ is set to be 8, and the dimension of each head $d_{k}$ equals 64. The dimension of the linear mapping layer is 512.

\section{Results and Analysis}
In this section, we will present extensive experimental results to evaluate the effectiveness of the proposed method.
\subsection{Evaluation of the proposed EW2}
\textbf{Comparison methods: }The Baseline in~\cite{9414027} utilizes the Deepspeech2 model  \cite{amodei2016deep} for training on the LibriSpeech train-clean-100 dataset with a CTC objective function and evaluates on different test sets.
For completeness, the time-domain DEMUCS as a front-end SE step in~\cite{9414027} will be compared, where the enhanced speech is directly used for ASR.
The AvT method utilized in~\cite{9414027} introduces a gradient reversal layer in prior to the model classification layer, such that the learned speech representations can be noise-invariant. In addition, the proposed EW2 model will also be compared with the original wav2vec2.0 method~\cite{NEURIPS2020_92d1e1eb}. Note that different combinations of pre-training and fine-tuning branches in Fig.~3 will be considered in experiments.

Table~\ref{tab:enhtable1} shows the ASR performances in terms of WER of the aforementioned approaches using the type-A (relatively stationary) and type-B (non-stationary) noises under different SNR conditions. The average performance is obtained by averaging the WERs over all input SNRs and noise types. From Table~\ref{tab:enhtable1}, it can be seen that although the structure of the proposed model is different from~\cite{9414027}, the proposed baseline system (i.e., wav2vec2.0 no pre-train clean fine-tune) achieves a comparable performance as compared to the baseline in~\cite{9414027}.  For wav2vec2.0, comparing `no pre-train clean fine-tune' and `no pre-train FreeSound fine-tune', it is clear that fine-tuning on noisy datasets can  improve the ASR performance under most noise conditions, that is, the noise robustness can be improved. As the combination of `clean pre-train FreeSound fine-tune' obtains a much better performance than `no pre-train FreeSound fine-tune' in both noisy and clean environments, the inclusion of a pre-training phase is rather beneficial for the robustness of ASR models. Comparing `clean pre-train FreeSound fine-tune' and `FreeSound pre-train FreeSound fine-tune' (the latter performs better), we find that the wav2vec2.0 model can still learn a robust speech representation under noisy scenarios.  Compared to DEMUCS or AvT, although wav2vec2.0 (i.e., FreeSound pre-train FreeSound fine-tune) can improve the performance on the test set under various noisy conditions, the performance on the clean test set drops significantly.

In order to see whether the wav2vec2.0 model is robust to noise types, we use the NoiseX-92 noise dataset to dynamically add noise to the train-clean-100 subset to obtain a noisy dataset for pre-training and then perform fine-tuning on noisy data. From Table~\ref{tab:enhtable1}, we can see that the choice of `NoiseX-92 pre-train FreedSound fine-tune' for wav2vec2.0 is better than the  `no pre-train FreedSound fine-tune' counterpart, indicating that the representations obtained by pre-training on other types of noisy data still have a good robustness. However, as the choice of `FreeSound pre-train FreeSound fine-tune'  leads to a decrease in WER compared to `NoiseX-92 pre-train FreeSound fine-tune', the data sources for pre-training and fine-tuning affects the performance of wav2vec2.0. That is, the noise data for pre-training and fine-tuning originating from different domains might degrade the ASR performance. This phenomenon also applies to the results of EW2.

More importantly, from Table~\ref{tab:enhtable1} we can see that the combination of `FreedSound pre-train FreedSound fine-tune' for the proposed EW2 method  is better than the same choice for wav2vec2.0 under both noisy and clean conditions. Using the clean speech as the pre-training targets can improve the performance on the noisy test set, and it is also ensured that the performance on the clean test set is not significantly degraded. In addition, the proposed EW2 method with `NoiseX-92 pre-train FreedSound fine-tune' outperforms the wav2vec2.0 counterpart, indicating that a better robustness against different noise types is obtained. Besides, although the proposed EW2 approach works slightly worse than DEMUCS on the clean test set, the performance under more-commonly noisy conditions is much better.

\begin{table}[]
\caption{ The average WER under different SNRs w/o the VQ module using clean or noisy targets in the pre-training stage.}
\label{tab:table1}
\centering
\begin{tabular}{l|c|c|c}
\hline
\textbf{Method}               & \textbf{Target type}            & \begin{tabular}[c]{@{}c@{}}\textbf{Vector}\\ \textbf{quantization}\end{tabular} & \begin{tabular}[c]{@{}c@{}}\textbf{Average WER} \\ \textbf{(SNR $\in$ 0$\sim$20 dB)}\end{tabular} \\ \hline
\multirow{2}{*}{EW2} & \multirow{2}{*}{Clean} & \cmark                                                         & 27.82                                                                        \\
                     &                        & \xmark                                                         & 28.70                                                                        \\ \hline
\multirow{2}{*}{EW2} & \multirow{2}{*}{Noisy} & \cmark                                                         & 33.06                                                                        \\
                     &                        & \xmark                                                         & 34.46                                                                        \\ \hline
\end{tabular}
\end{table}

\subsection{Evaluation on the necessity of VQ and consistency loss }
In order to better understand the function of the VQ module, we conduct comparative experiments using clean and noisy targets, and the results are shown in Table~\ref{tab:table1}. It is obvious that for both target types in the pre-training stage, the performance of using the VQ module can be improved compared to the case of without VQ, indicating the necessity of the VQ module in noisy scenes. As the WER reduction of using clean targets is larger than using noisy ones, the clean targets in the pre-training stage are beneficial for speech representation learning.

\begin{table}[]
\caption{ The average WER under different SNRs w/o the consistency loss in the pre-training stage.}
\label{tab:table4}
\centering
\begin{tabular}{l|c|c}
\hline
\textbf{Method} & \begin{tabular}[c]{@{}c@{}}\textbf{Consistency}\\ \textbf{loss}\end{tabular} & \textbf{Average WER (SNR $\in$ 0$\sim$20 dB)} \\ \hline
EW2    & \cmark                                                      & 27.82                             \\
EW2    & \xmark                                                      & 29.23                             \\ \hline
\end{tabular}
\end{table}

\begin{table*}[!t]
\caption{ The performance comparison of different methods on type-A noise test sets at different SNRs when pre-train on 460 hours librispeech data and fine-tune on 100 hours librispeech data. SE stands for speech enhancement module and FE stands for feature encoder module in Fig.~\ref{fig:robustse}.}
\label{tab:table5}
\centering
\begin{tabular}{l|c|c|c|ccccc}
\hline
\multirow{3}{*}{\textbf{Method}} & \multirow{3}{*}{\begin{tabular}[c]{@{}c@{}}\textbf{Initialize with}\\ \textbf{pre-trained model}\end{tabular}} & \multirow{3}{*}{\begin{tabular}[c]{@{}c@{}}\textbf{Update}\\ \textbf{SE}\end{tabular}} & \multirow{3}{*}{\begin{tabular}[c]{@{}c@{}}\textbf{Update}\\ \textbf{FE}\end{tabular}} & \multicolumn{5}{c}{\textbf{WER under SNR}}                                                                                                                               \\ \cline{5-9}
                        &                                                                                             &                                                                        &                                                                        & \multicolumn{1}{c|}{\textbf{Babble}}       & \multicolumn{1}{c|}{\textbf{Airport/Station}} & \multicolumn{1}{c|}{\textbf{AC/Vacuum}}    & \multicolumn{1}{c|}{\textbf{Cafe}}         & \textbf{Average}      \\ \cline{5-9}
                        &                                                                                             &                                                                        &                                                                        & \multicolumn{1}{c|}{\textbf{0$\sim$20 dB}} & \multicolumn{1}{c|}{\textbf{0$\sim$20 dB}}    & \multicolumn{1}{c|}{\textbf{0$\sim$20 dB}} & \multicolumn{1}{c|}{\textbf{0$\sim$20 dB}} & \textbf{0$\sim$20 dB} \\ \hline
EW2                      & \cmark                                                                                       & -                                                                      & \cmark                                                                  & \multicolumn{1}{c|}{20.96}        & \multicolumn{1}{c|}{16.06}           & \multicolumn{1}{c|}{15.74}        & \multicolumn{1}{c|}{12.16}        & 16.23        \\
SEW2                     & \xmark                                                                                       & \cmark                                                                  & \cmark                                                                  & \multicolumn{1}{c|}{35.12}        & \multicolumn{1}{c|}{29.94}           & \multicolumn{1}{c|}{29.16}        & \multicolumn{1}{c|}{23.22}        & 29.36        \\
SEW2                     & \cmark                                                                                       & \cmark                                                                  & \xmark                                                                  & \multicolumn{1}{c|}{20.04}        & \multicolumn{1}{c|}{15.20}           & \multicolumn{1}{c|}{14.96}        & \multicolumn{1}{c|}{11.56}        & 15.44        \\
SEW2                     & \cmark                                                                                       & \cmark                                                                  & \cmark                                                                  & \multicolumn{1}{c|}{19.64}        & \multicolumn{1}{c|}{14.92}           & \multicolumn{1}{c|}{14.68}        & \multicolumn{1}{c|}{11.42}        & 15.17        \\
EW2 + SEW2                  & \cmark                                                                                       & \cmark                                                                  & \cmark                                                                  & \multicolumn{1}{c|}{\textbf{18.63}}        & \multicolumn{1}{c|}{\textbf{14.36}}           & \multicolumn{1}{c|}{\textbf{14.06}}        & \multicolumn{1}{c|}{\textbf{11.16}}        & \textbf{14.55}        \\
EW2 + SEW2                  & \xmark                                                                                       & \cmark                                                                  & \cmark                                                                  & \multicolumn{1}{c|}{19.56}        & \multicolumn{1}{c|}{14.59}           & \multicolumn{1}{c|}{14.43}        & \multicolumn{1}{c|}{11.32}        & 14.98        \\ \hline
\end{tabular}
\end{table*}

\begin{figure}[!t]
  \centering
  \includegraphics[width=0.45\textwidth]{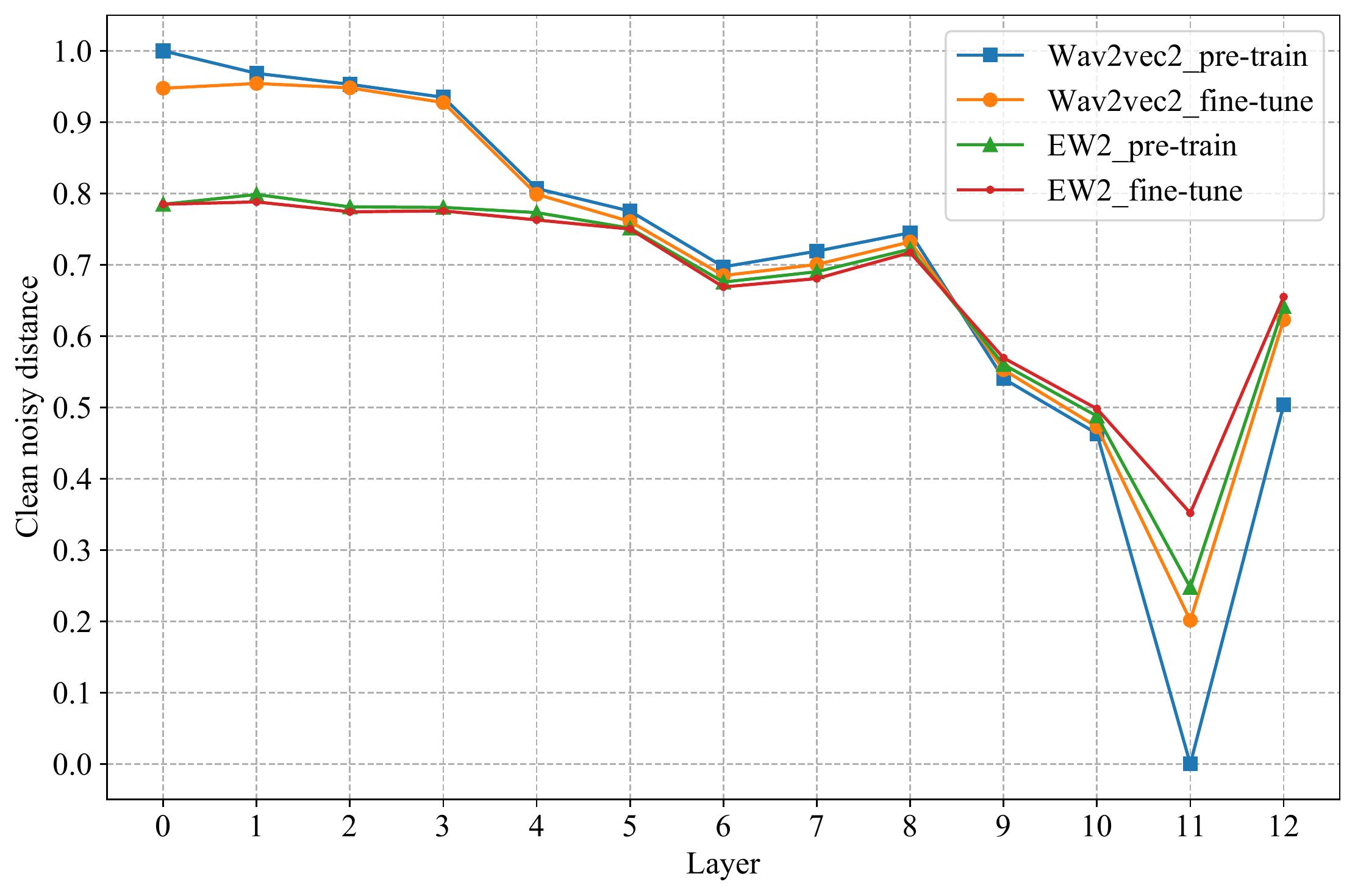}
  \caption{The normalized $\ell_2$ distance between clean and noisy representations at different layers, where layer 0 represents the feature encoder.}
  \label{fig:cndistance}
\end{figure}

We further compare the performance of the proposed EW2 model  with or without the consistency loss function in Table~\ref{tab:table4}. It is clearly shown that introducing a consistency loss function in the output of the feature encoder can improve the ASR performance. In order to show the necessity of the consistency loss, we visualize the normalized $\ell_2$ distance between clean and noisy features at different layers in Fig.~\ref{fig:cndistance}. The distance between low-level clean features and noisy features is larger, and high-level clean features and noisy features have smaller distances, indicating that low-level features are more susceptible to noise interference. Therefore, adding a consistency loss function to low-level features can reduce the distance between clean features and noisy features and enhance the noise robustness of ASR models.

\subsection{Evaluation on joint training of SE and EW2}

We evaluate the effectiveness of the proposed joint training of SE and EW2 on type-A and type-B noise test sets, which are shown at the bottom of Table~\ref{tab:enhtable1}.
For SEW2, comparing `no pre-train FreeSound fine-tune' and `FreeSound pre-train FreeSound fine-tune' it is clear that introducing an additional pre-training stage can improve the performance. However, comparing `FreeSound pre-train FreeSound fine-tune' of SEW2 and `FreeSound pre-train FreeSound fine-tune' of EW2, the performance of SEW2 is worse than that of EW2, that is, only feeding the noisy speech to the SE module together with a random initialization for pre-training cannot guarantee a robust speech representation. This is due to the fact that the SE operation would cause the speech distortion problem, which deteriorates the representation quality of the model in the pre-training stage.
To alleviate this problem, we adopt the dual-attention fusion module in Section III-C to fuse the noisy feature and enhanced feature in the pre-training stage, such that the fused context information can be learned.
Under the `no pre-train FreeSound fine-tune' condition, comparing EW2 + SEW2 and SEW2 methods we find that in case the model is directly fine-tuned with random initialization even without pre-training, the fusion of noisy features and enhanced features can improve the performance.
Under the `FreeSound pre-train FreeSound fine-tune' condition, comparing EW2, SEW2 and EW2 + SEW2 methods, it is clear that the latter performs the best, meaning that fusing the noisy  and enhanced features in the pre-training stage can alleviate the speech distortion problem and thus reduce the WER.
Comparing EW2+SEW2 (concat) and EW2+SEW2, the latter clearly performs better in the cases of  both `no pre-train FreeSound fine-tune' and `FreeSound pre-train FreeSound fine-tune'.  This implies that the proposed dual attention fusion method outperforms the classic concatenation technique for feature fusion.

To verify the effectiveness of the proposed method on a larger dataset, we use 460 hours of LibriSpeech clean-noisy data to continue pre-train our model initialized with the public pre-trained wav2vec2.0 model for 50k updates, which is  then fine-tuned on the noisy data. The experimental results on type-B noise test sets are shown in Table~\ref{tab:table5}.
In the case of SEW2, the joint training of the SE module and the self-supervised model with random initialization cannot obtain a good representation. In case a pre-trained self-supervised model is utilized for initialization, no matter whether the feature encoder is updated or not, the performance is much better than the case without a pre-trained model as initialization, indicating that an off-the-shelf pre-trained model can potentially decrease the impact of speech distortion to a certain extent.
Besides, as the entire model is updated with a small learning rate in the fine-tuning stage, the front-end SE module is adjusted in line with the back-end ASR, leading to a  performance gain.
Comparing the EW2, SEW2 and EW2+SEW2 methods, EW2+SEW2 can achieve a very promising WER regardless of the inclusion of a pre-trained model as initialization. That is, the proposed joint training method for SE and self-supervised speech representation learning is robust against model setup and noise conditions.

\subsection{Analysis of the importance of initialization}
\begin{figure}[!t]
  \centering
  \includegraphics[width=0.4\textwidth]{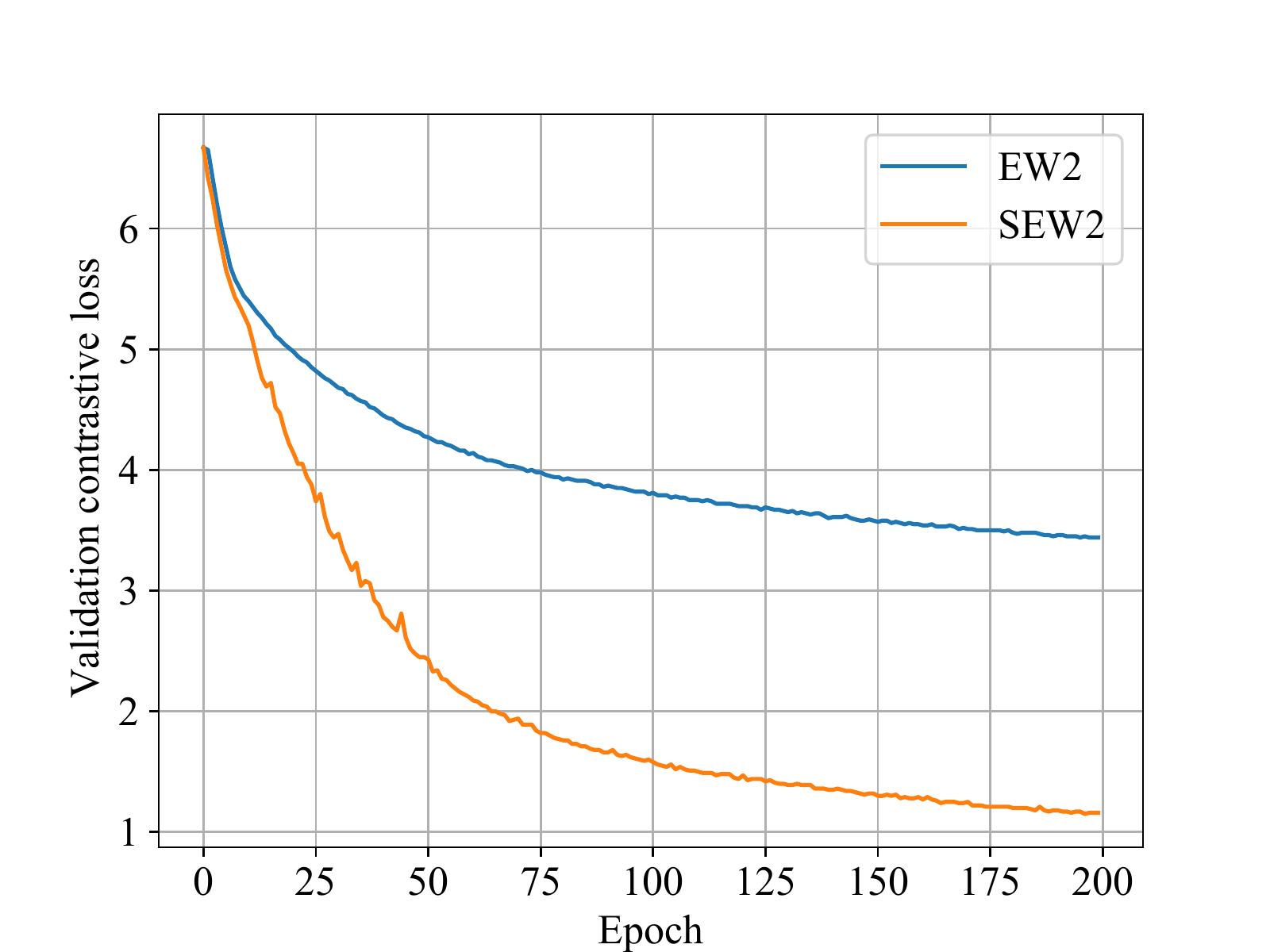}
  \caption{The validation contrastive loss versus the number of training epochs in the pre-training stage.}
  \label{fig:validloss}
\end{figure}

In order to further investigate the effect of the random initialization for the SE module and the self-supervised model (e.g., SEW2 method in Table~\ref{tab:table5}) on the speech representation learning, we visualize the loss function on the validation set versus the number of training epochs in the pre-training stage in Fig.~\ref{fig:validloss}. It can be seen that with the addition of the SE module, the loss on the validation set in the pre-training stage drops rapidly. This reveals that the SE operation would cause speech distortion and smooth the detailed information contained in speech, and  the contextual representation becomes easier to be leveraged for the prediction of the masked information in the pre-training stage, which, however degrades the quality or fidelity of the learned speech representation.

\begin{figure}[!t]
  \centering
  \includegraphics[width=0.4\textwidth]{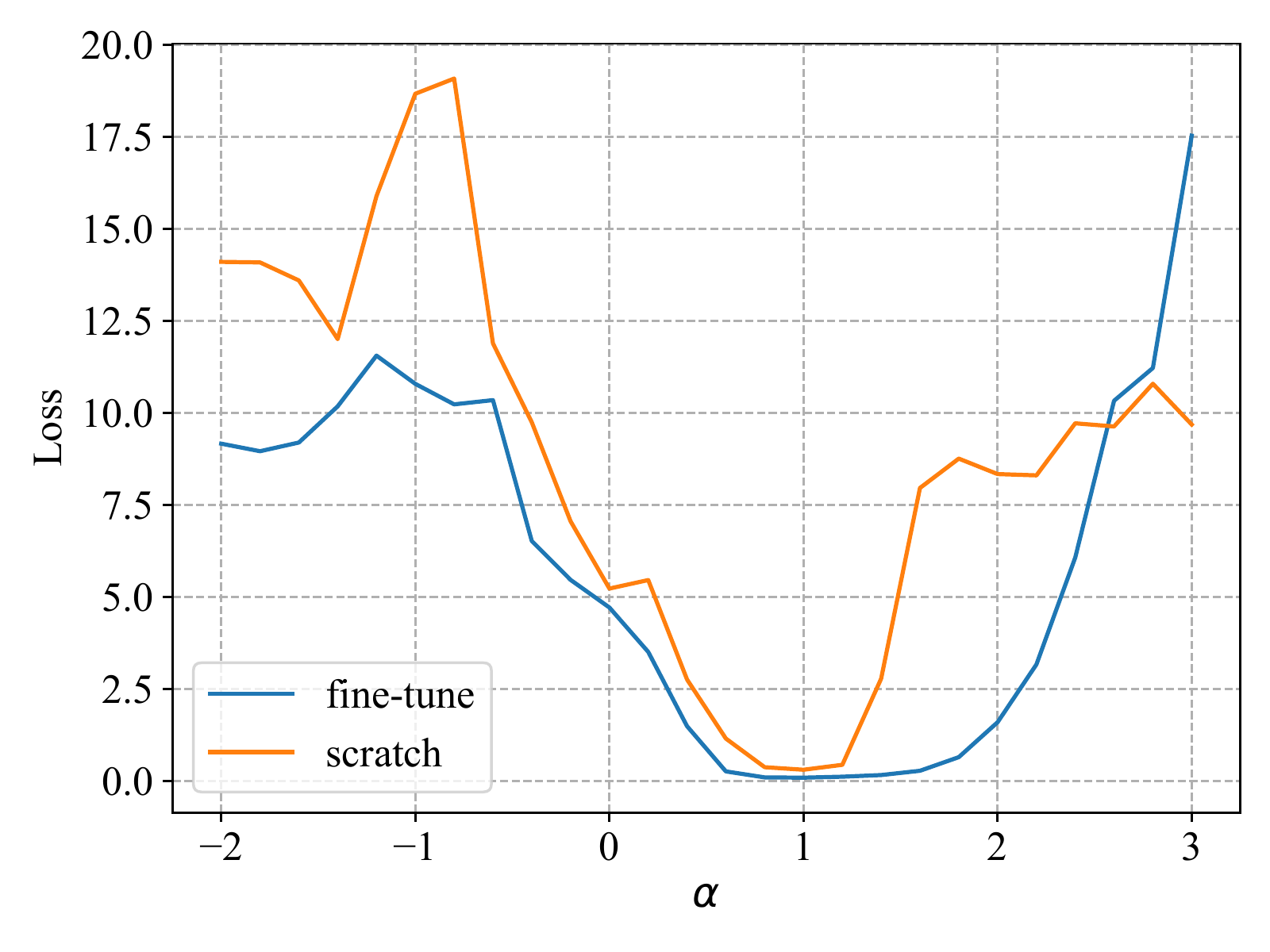}
  \caption{The 1D loss curves, where the blue and orange curves represent fine-tuning EW2 and training from scratch, respectively.}
  \label{fig:lossfinetune}
\end{figure}

\begin{figure}[!t]
  \centering
  \includegraphics[width=0.48\textwidth]{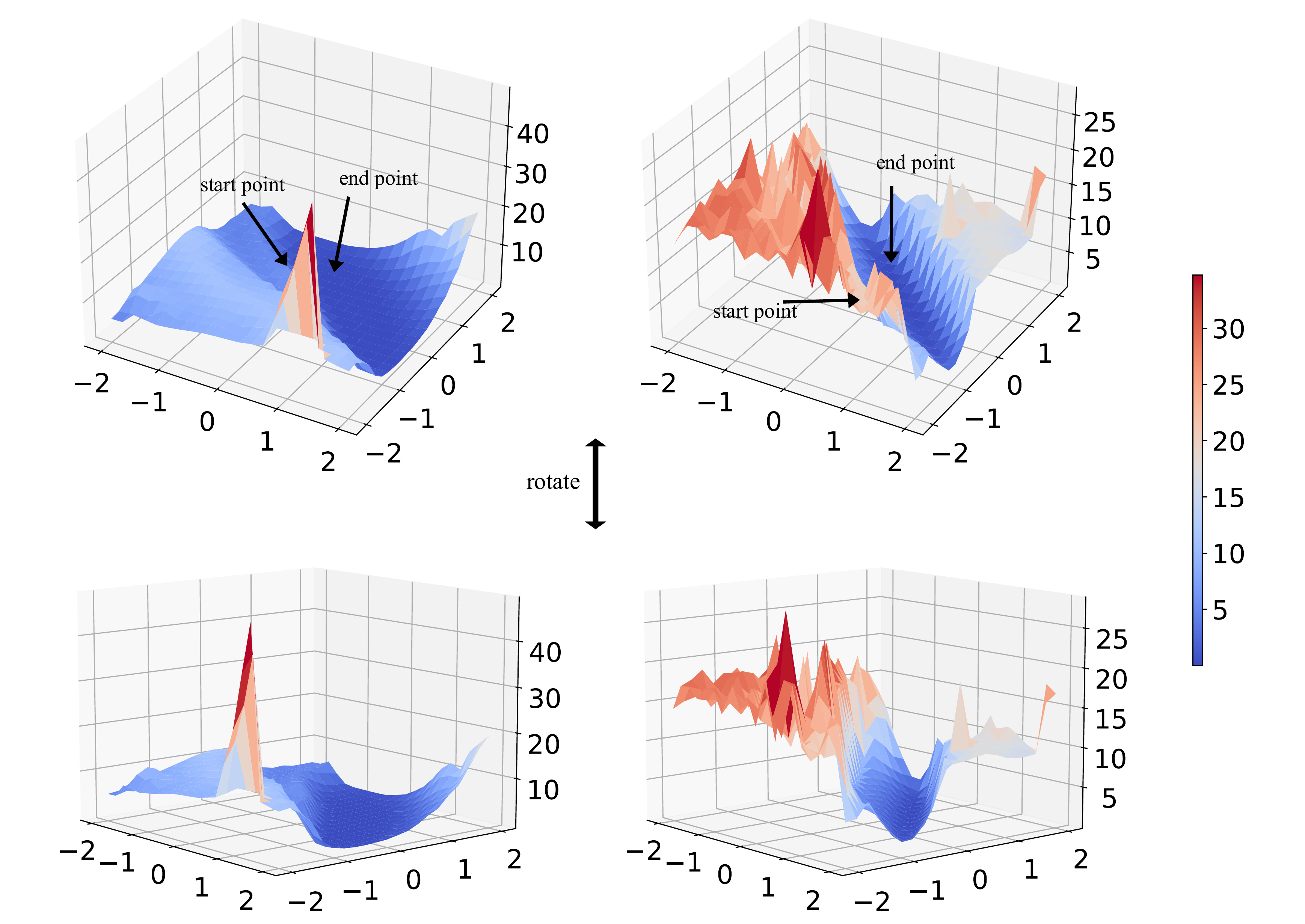}
  \caption{The 2D loss surfaces of fine-tuning EW2 (left) and training from scratch (right), where the bottom row is obtained via rotating.}
  \label{fig:losslandscape1}
\end{figure}

We further analyze the effect of the application of an off-the-shelf pre-trained self-supervised model for initialization on the robust speech representation learning (e.g., SEW2 method in Table~\ref{tab:table5}). In principal, a pre-trained self-supervised model has potential to enable a good initial point, which is rather conducive for the model optimization to converge.
To show this, we follow the method in~\cite{li2018visualizing,hao2019visualizing} and visualize the one-dimensional (1D) and two-dimensional (2D) loss landscape of the model that is fine-tuned on the 100 hours LibriSpeech data, which can be initialized either randomly or using an existing pre-trained model, so that we can compare the difference between two learning paradigms. Let $\boldsymbol{\theta}_0$ represent the initial parameters, which can then denote the pre-trained parameters of fine-tuning EW2 or the randomly initialized parameters of training from scratch. Let $\boldsymbol{\theta}_1$ denote the model parameters after fine-tuning. The 1D loss curve function $f(m)$ shows the loss value along the searching direction, which is defined as
\[
f(m)=J(\boldsymbol{\theta}_0 + m \boldsymbol{\delta}_1),\]
 where $m$ is a scalar, $\boldsymbol{\delta}_1 = \boldsymbol{\theta}_1 - \boldsymbol{\theta}_0$ is the optimization direction, and $J(\boldsymbol{\theta})$ is the loss function given $\boldsymbol{\theta}$. For the brevity of visualization, we set $m\in$ [-2, 3] and discretize the 1D loss curve. Similarly, the 2D loss surface function is defined as
\[f(m,n)=J(\boldsymbol{\theta}_0 + m \boldsymbol{\delta}_1 + n \boldsymbol{\delta}_2),\]
 where $m,n$ are scalars. One optimization direction is $\boldsymbol{\delta}_1 = \boldsymbol{\theta}_1 - \boldsymbol{\theta}_0$ and the other is $\boldsymbol{\delta}_2 = \boldsymbol{\theta}_2 - \boldsymbol{\theta}_0$, where $\boldsymbol{\theta}_2$ represents the fine-tuned parameters on another dataset. We set the range of both $m$ and $n$ to be [-2, 2] and sample 21 points at each axis. It is clear that in case $m=0$ and $n=0$, $\boldsymbol{\theta}_0$ denotes the initial point,  and in case  $m=1$ and $n=0$, $\boldsymbol{\theta}_0+\boldsymbol{\delta_1}$ is the ending point of model optimization.
The 1D and 2D loss curves are shown in Fig.~\ref{fig:lossfinetune} and Fig.~\ref{fig:losslandscape1}, respectively. It can be seen that fine-tuning with a pre-trained model has a wider optimal area than fine-tuning with a random initialization model, which means that in case of using a pre-trained model, the slight turbulence of the model parameters can not seriously affect the model performance.
In addition, fine-tuning with a pre-trained model results in a smoother loss surface near the starting point than fine-tuning with random initialization, which indicates that the pre-trained model provides a better initialization point and it becomes easier and faster for the model to converge to the optimal status. Therefore, when using a pre-trained self-supervised model, adding an extra SE module can easily find the optimum and thus improve the model performance.

\begin{table}[!t]
\caption{ The performance comparison of different methods on CHiME-4 one-channel dataset.}
\label{tab:table6}
\centering
\begin{tabular}{lccc}
\hline
\multirow{2}{*}{\textbf{Model}}                      & \multirow{2}{*}{\textbf{LM}} & \multicolumn{2}{c}{\textbf{WER}} \\ \cline{3-4}
                                            &                     & \textbf{dt05\_real} & \textbf{et05\_real} \\ \hline
\hline
\multicolumn{4}{l}{\textbf{Supervised}}                                                              \\
DNN baseline~\cite{vincent20164th}                                & N-gram              & 11.6       & 23.7       \\
Du et al.~\cite{du2016ustc}                                   & LSTM                & 4.5        & 9.2        \\
Menne et al.~\cite{menne2016rwth}                                & LSTM                & 5.1        & 9.3        \\
Wang et al.~\cite{9103053}                                 & LSTM                & 3.5        & 6.8        \\ \hline
\hline
\multicolumn{4}{l}{\textbf{Self-supervised}}                                                         \\
Wang et al.(960h)~\cite{wang2021improving}                           & LSTM                & 5.0        & 9.0        \\
Wang et al.(60kh)~\cite{wang2021improving}                           & LSTM                & 2.8        & 5.8        \\
\hline
Wav2vec2.0 Base~\cite{gao2021data}                         & None                & 10.3       & 17.8       \\
Gao et al.~\cite{gao2021data}                                  & None                & 8.7        & 15.8       \\
\hline
\multirow{2}{*}{Wav2vec2.0 Base~\cite{wang2021wav2vec}}        & None                & 10.6       & 17.6       \\
                                            & LSTM                & 3.7        & 7.2        \\
\multirow{2}{*}{Wav2vec-switch~\cite{wang2021wav2vec}}             & None                & 10.0       & 16.5       \\
                                            & LSTM                & 3.5        & 6.6        \\
\hline
\multirow{2}{*}{HUBERT Base~\cite{9585401}}                     & None                & 10.4       & 17.0       \\
                                            & LSTM                & 3.8        & 7.1        \\
\hline
\multirow{2}{*}{Wav2vec2.0 Base (Ours)~\cite{zhu2022noise}} & None                & 10.5       & 17.3       \\
                                            & Transformer         & 3.8        & 7.5        \\
\multirow{2}{*}{EW2 (Ours)~\cite{zhu2022noise}}                 & None                & 9.4        & 15.6       \\
                                            & Transformer         & 3.5        & 6.4        \\ \hline
\hline
\multicolumn{4}{l}{\textbf{Speech enhancement + self-supervised}}                                    \\
Chang et al.~\cite{chang2022end}                                & Transformer         & 2.03       & 3.92       \\
\hline
\multirow{2}{*}{EW2 + SEW (Ours)}           & None                & 8.2        & 14.3       \\
                                            & Transformer         & 3.0        & 5.9        \\ \hline
\end{tabular}
\end{table}

\subsection{Performance evaluation on the CHiME-4 dataset}
Finally, we validate the proposed method in realistic noisy scenes using  the CHiME-4 dataset. The experimental results are shown in Table~\ref{tab:table6}. The one-channel real-world noisy data `dt05\_real' and `et05\_real' are utilized for model validation and evaluation, respectively. For comparison, we also include the results of some supervised and self-supervised methods. The top part of Table~\ref{tab:table6} shows the representative state-of-the-art results on single-channel data using supervised methods, most of which employ extensive data augmentation and auxiliary data processing techniques, e.g., speaker adaptation and model fusion. The middle part shows the representative self-supervised pre-training methods on the single-channel data. Self-supervised pre-training methods utilize a large amount of unlabeled data for pre-training and are fine-tuned on a much smaller amount of labeled dataset, which do not require complex data processing operations. For example, Wang et al.~\cite{wang2021improving} utilized a large pre-trained model to achieve a WER of 2.8 on the dt05\_real validation set and a WER of 5.8 on the et05\_real test set, which is even better than the best supervised method~\cite{9103053} with a WER of 3.5 on the dt05\_real validation set and a WER of 6.8 on the et05\_real test set. For the self-supervised pre-training methods, without an LM, the proposed EW2~\cite{zhu2022noise} achieves a WER of 9.4 on the validation set and a WER of 15.6 on the test set, which performs better than wav2vec-swith with a WER of 10.0 on the validation set and a WER of 16.5 on the test set. With the inclusion of LM models, their performance becomes comparable. For the proposed joint training method without LM, the WER on the validation set is 8.2 and the WER on the test set is 14.3, which is about 10\% relative improvement compared to EW2. Applying a Transformer-based LM, the performance gain of the proposed method is consistently achieved. These also show that the LM is rather beneficial for improving the ASR performance. It is worth mentioning that to the best of our knowledge Chang et al.~\cite{chang2022end} achieves the best result on the monaural CHiME-4 ASR task so far with a WER of 2.03 on the validation set and a WER of 3.92 on the test set.
In~\cite{chang2022end},  the self-supervised pre-trained WavLM  model is taken as the feature extractor, the trained SE module and the trained ASR model are fine-tuned in a cascaded fashion, and all modules are pre-trained separately. This independent pre-training and fine-tuning would have a serious performance drop in case the model is not carefully initialized. For example, as shown in~\cite{chang2022end} in the case of random initialization, the performance or the convergence cannot be guaranteed, due to the fact that the model depth is too large and the gradient back propagation is blocked. As the source code of~\cite{chang2022end} has not been published, the implementation details are still unknown. As in practice random initialization is more promising for training and easier for implementation, the proposed method reaches the state-of-the-art performance from this perspective.

\section{Conclusion}
In this paper, we investigated the  jointly pre-training of the SE and the self-supervised model for noise-robust ASR, in which the original noisy waveform or the waveform after SE is fed into the self-supervised model to learn the contextual representation and the quantified clean speech provides targets for the pre-training model. The deployment of noisy, enhancement and clean branches allows for several noise-robust ASR variants, which turns out the generality of the proposed joint training approach. We also proposed a dual-attention module for the feature fusion. It was shown that both the SE and the self-supervised model can improve the ASR performance in noisy scenes.
Besides, the dual-attention fusion module can compensate the information loss in separately using SE or pre-training models and thus reduce the distortion caused by SE to a certain extent. The inclusion of VQ operations and a consistency loss between clean and noisy features is also important for improving the noise robustness in practical noisy environments. Although the proposed method can be initilized randomly, we found that using a pre-trained model for  initialization can further reduce the speech distortion and reduce the WER, which was verified based on the optimization theory. Compared to~\cite{chang2022end}, which requires a careful model initialization, the proposed method exhibits a more promising robustness against random initialization as the sacrifice in performance is quite small.  This is rather important for the implementation of noise-robust ASR systems in practice.
In the future, we will consider the joint training for speech intelligibility enhancement and self-supervised pre-training models, which might potentially reduce the perceptual instrumental speech distortion compared to the MSE-based SE models, as the speech intelligibility is more related to the ASR capability.

\bibliographystyle{IEEEtran}
\bibliography{strings,refs}

\end{document}